\documentclass[]{interact}

\usepackage{epstopdf}
\usepackage[caption=false]{subfig}

\usepackage[numbers,sort&compress]{natbib}
\bibpunct[, ]{[}{]}{,}{n}{,}{,}
\renewcommand\bibfont{\fontsize{10}{12}\selectfont}

\theoremstyle{plain}

\theoremstyle{definition}

\theoremstyle{remark}

\begin{document}


\title{Random Matrix Time Series}

\author{
	\name{Peiyuan Teng\textsuperscript{a,b,*}, Min Xu\textsuperscript{a}}
	\affil{\textsuperscript{a} The Ohio State University, 191 West Woodruff Ave,
		Columbus, OH 43210, USA; \textsuperscript{b} U.S. Bancorp, 800 Nicollet Mall, Minneapolis, MN, 55402, USA;}
	\email{\textsuperscript{*}teng.73@osu.edu}
}

\maketitle
\begin{abstract}
	In this paper, a time series model with coefficients that take values from random matrix ensembles is proposed. Formal definitions, theoretical solutions, and statistical properties are derived. Estimation and forecast methodologies for random matrix time series are discussed with examples. Random matrix differential equations and potential applications of the time series model are suggested at the end.\\
	
\end{abstract}


\section{Introduction.}
Random Matrix theories were first introduced in statistics by Wishart in 1928\cite{wishart1928generalised}. In 1951, they were introduced into nuclear physics by Wigner  \cite{wigner1951statistical}. Since then, Random Matrix theories have become a rich research topic in both the Mathematics and Physics communities. In 1962, Dyson introduced the classification of random matrix ensembles based on symmetries\cite{dyson1962statistical}, which are the Gaussian unitary ensemble (GUE), the Gaussian orthogonal ensemble (GOE), and the Gaussian symplectic ensemble (GSE). Various applications can be found in the research of statistics, integrable systems, operator algebras, number theory, or in statistical physics, quantum chromodynamics, and quantum gravity\cite{forrester2003developments}.\\

One of the most commonly used random matrix ensembles, the Gaussian orthogonal ensemble (GOE), is described by the density
\begin{equation}
	\dfrac {1}{Z_{GOE(n)}}  e^{-{\frac {n}{4}} tr H^{2}},
\end{equation}
on the space of $n \times n$ real symmetric matrices $H$.
$Z_{GOE(n)}$ is a normalization constant. \\

The Gaussian unitary ensemble (GUE) is described by the density

\begin{equation}
	\dfrac {1}{Z_{GUE(n)} }e^{-{\frac {n}{2}} {tr} H^{2}},
\end{equation}

on the space of $n \times n$ Hermitian matrices $H$.  $Z_{GUE(n)}$ is a normalization constant. \\

In statistics, time series analysis refers to methods for analyzing time-series data.  Statistics and other characteristics can be acquired by fitting a time series model with actual data. Those fitted characteristics or time series model parameters can also be used to generate forecasts. Typical time series models are autoregressive (AR) models, vector autoregression (VAR),  autoregressive integrated moving average (ARIMA) models, and generalized autoregressive conditional heteroskedasticity (GARCH) models.  Readers can refer to Ref.\cite{box2015time} for a detailed review of those models.  \\

The vector autoregression (VAR) models, as an example, can be viewed as another starting point for this paper. A p-th order VAR, denoted as VAR(p), can be written as.
\begin{equation}\label{Var}
	X_{t}=c+A_{1}X_{t-1}+A_{2}X_{t-2}+\cdots +A_{p}X_{t-p}+e_{t}.
\end{equation}

The variables  $X_{t}$ is a k-th order vector. The variable  $X_{t-i}$ is the i-th lag of $X_t$. The variable c is the intercept of the model. $A_p$ are time-invariant $k \times k $ matrices and $e_t$ is the error term.\\

Since both Random Matrix theory and Time-series theory are very useful statistical methods, it is natural to consider combining those two theories. Several pioneering works have been published in this direction. For example, In Ref. \cite{plerou1999universal}, the authors use methods of random matrix theory to analyze the cross-correlation matrix of stock price changes and find that the cross-correlation matrix has the universal properties of the Gaussian orthogonal ensemble of random matrices. Subsequent research can be found at \cite{biely2008random}\cite{laloux2000random}\cite{ormerod2008random}.  In Ref. \cite{fossion2013random}, the spectra of random matrices treated as time series are researched.  Ref. \cite{ye2017distinguishing} proposes a random matrix theory (RMT) approach to distinguishing chaotic time series from noise.

In this paper, those two theories are combined from a new perspective. Time series vectors are defined iteratively with iteration coefficients taking values from random matrix ensembles. For simplicity, we call our time series models: n-th order time series with random matrix coefficients or RMTS(n). A formal definition will be given, and from its definition, theoretical solutions and statistical properties such as expectation and variance will be derived. Furthermore, we extend those derivations to complex random matrices. Estimation and forecast methodologies for random matrix time series will also be discussed with examples. For comparison, the first-order random matrix differential equation is defined in parallel to the random matrix time series. Potential applications will also be suggested at the end.\\

This paper is organized as follows: In Section 2, first-order random matrix time series models, RMTS(1) are researched. In Section 3, n-th order models, RMTS(n), are discussed. In Section 4, the first-order random matrix differential equation is proposed.  Section 5 suggests some potential applications.

\section{First order time series with random matrix coefficients, RMTS(1).}

The first-order time series with random matrix coefficients or RMTS(1) can be defined similarly as VAR(1)(see Equation \ref{Var}). The difference is that the error term is no longer needed. Instead, the matrix coefficient, which is the random matrix, contains not only the estimation parameters but also absorbs randomness.\\

\textbf{Defination}: A first order vector autoregressive time series with random matrix coefficients, denoted by RMTS(1), is defined by the iterative relation:
\begin{equation} \label{eq:RMTS1}
	\mathbf {X} (T+1)=\mathbf{A}(T)\mathbf {X} (T)+\mathbf {b}(T).
\end{equation}
In this equation, $\mathbf {X} (T)$ is the vector time series, $\mathbf{A}(T)$ is a random matrix at time $T$. $\mathbf{b}(T)$ is a random vector whose elements are independent Gaussian random variables. $\mathbf{A}(T)$ can take value from a Gaussian ensemble, or more generally, a matrix whose elements are random numbers. The density distribution of  $\mathbf{A}(T)$ is assumed to be time-invariant.\\

The readers might notice that, if $\mathbf{A}(T)$ and $\mathbf{b}(T)$ are fixed numbers rather than random numbers, equation \ref{eq:RMTS1} is essentially a difference equation. Due to this similarity, many methods in the difference equation can be similarly extended to the study of this type of times series. Readers can refer to Ref. \cite{agarwal2000difference}  for a comprehensive discussion of difference equations.\\

\subsection{Solutions of RMTS(1)}

In this part, we'd like to derive an analytical solution for Equation \ref{eq:RMTS1}. We can start with a homogeneous RMTS(1) with no constant term.

Consider the case when $\mathbf {b} (T)=0$, we have equation: 
\begin{equation*}
	\mathbf {X}(T+1)=\mathbf{A}(T)\mathbf {X} (T).
\end{equation*}
Define a product of random matrices by

\begin{equation}
	\mathbf{V}(T)=\prod_{k=0}^{T}\mathbf{A}(k).
\end{equation}
Therefore, by utilizing the iteration relation, we can easily get

\begin{equation*}
	\mathbf {X} (T+1)=\mathbf{V}(T)\mathbf {X} (0).
\end{equation*}\\

Now let's consider a generic case where $\mathbf {b} (T)\ne0$, assuming $V(T)$ has an inverse for every $T$ (which is true in general due to the randomness of $\mathbf{A}(T)$), define $\mathbf {C} (T)$ by the equation (with $\mathbf {C} (0)=\mathbf {X} (0)$).

\begin{equation}
	\mathbf {X} (T+1)=\mathbf{V}(T+1)\mathbf {C} (T+1).
\end{equation}

Plugging this equation into Equation \ref{eq:RMTS1}, and we have

\begin{align*} 
	\mathbf{V}(T+1)\mathbf {C} (T+1) &=	\mathbf {X} (T+1) \\&=\mathbf{A}(T)\mathbf {X} (T)+\mathbf {b} (T)\\ &=\mathbf{A}(T)\mathbf{V}(T)\mathbf {C} (T)+\mathbf {b} (T)\\ &=\mathbf{V}(T+1)\mathbf {C} (T)+\mathbf {b} (T),
\end{align*} 

which is 
\begin{equation*}
	\mathbf{V}(T+1)\mathbf {C} (T+1)=\mathbf{V}(T+1)\mathbf {C} (T)+\mathbf {b} (T),
\end{equation*}

Multiply  both sides of this equation by $\mathbf{V}(T+1)^{-1}$, we have

\begin{equation*}
	\mathbf {C} (T+1)-\mathbf {C} (T)=\mathbf{V}(T+1)^{-1}\mathbf {b} (T).
\end{equation*}

So 

\begin{equation*}
	\mathbf {C} (T+1)=\mathbf {C} (0)+\sum_{i=0}^{T} \mathbf{V}(i+1)^{-1}\mathbf {b} (i).
\end{equation*}

We can write the solutions of $\mathbf {X} (T+1)$ as

\begin{equation}\label{gs}
	\mathbf {X} (T+1)=\mathbf{V}(T+1)(\mathbf {C} (0)+\sum_{i=0}^{T} \mathbf{V}(i+1)^{-1}\mathbf {b} (i)),
\end{equation}

with a  matrix $\mathbf{V}(T)$ defined as
\begin{equation*}
	\mathbf{V}(T)=\prod_{k=0}^{T}\mathbf{A}(k).
\end{equation*}
\\

\textbf{Products of random matrices:} 

The variable $\mathbf{V}(T)$ is a product of random matrices. As an ongoing research direction in the Random Matrix theory, the study of the product of random matrices produces many valuable theoretical results and practical applications. From a theoretical perspective, existing research focused on the probability density of the products random matrices\cite{ipsen2015products}\cite{akemann2015recent}, the joint probability density of the eigenvalues\cite{ipsen2014weak}  or the singular values\cite{akemann2013products}. Those results were applied to matrix-valued diffusions\cite{gudowska2003infinite}, quantum chromodynamics\cite{osborn2004universal} and wireless telecommunication \cite{tulino2004random}, etc. Readers can refer to Ref. \cite{ipsen2015products}, in which an analytical representation of $\mathbf{V}(T)$ in the perspective of the product of random matrices was given. 

In Section 4, distributions of the product of random matrices will play a major role in discussing differential equation solutions.

\subsection{Maximum likelihood estimation for  RMTS(1)}

Generically, each element $a_{ij} $ of random matrix $\mathbf {A}$ is assumed to have a distribution $a_{ij} \sim N(r_{ij},\sigma_{ij}^2)$, and those distributions are time-invariant. Given actual data, it is important to derive the parameter estimation methodology.  \\

From Equation \ref{eq:RMTS1}, using the properties of the sum of Gaussian random variables, it is easy to see that the  i-th row of  $\mathbf{A}(T)\mathbf {X} (T)$ has a  distribution of $N(\sum_j(r_{ij}X_j(T)),\sum_j\sigma_{ij}^2X_j(T)^2)$. Assuming $b_i(T)\sim N(b_{i},\sigma_{b_i}^2)$,
Then the i-th row of  $\mathbf {X} (T+1)$ has a  distribution of

\begin{equation}
	N(\sum_j(r_{ij}X_j(T)+b_i),\sum_j\sigma_{ij}^2X_j(T)^2+\sigma_{b_i}^2). 
\end{equation}

Given the observations from period 0 to $T_f+1$ The likelihood function can be written as
\begin{multline}
	L(r_{ij},\sigma_{ij},b_{i},\sigma_{b_i}^2;\mathbf {X} (0)|\mathbf {X} (1),\mathbf {X} (1)|\mathbf {X} (2),...,\mathbf {X} (T_f)|\mathbf {X} (T_f+1))\\
	=\prod_{T=0}^{T_f+1}\prod_{i}\frac{1}{{\sqrt {2\pi (\sum_j\sigma_{ij}^2X_j(T)^2+\sigma_{b_i}^2) } }}e^{{{ - \left( {X_i(T+1) -\sum_j r_{ij}X_j(T)-b_i } \right)^2 } \mathord{\left/ {\vphantom {{ - \left( {X_i(T+1)-\sum_j r_{ij}X_j(T)-b_i } \right)^2 } {2(\sigma_{ij}^2X_j(T)^2+\sigma_{b_i}^2 )^2 }}} \right. \kern-\nulldelimiterspace} {2(\sum_j\sigma_{ij}^2X_j(T)^2+\sigma_{b_i}^2)}}}.
\end{multline}

Given $\mathbf {X} (T)$, we can find the best-estimated parameters by maximizing this function. Practically, if we have more than one time series(with the same parameter distribution) available, the likelihood function can be written correspondingly by multiplying the likelihood of all the available time series.\\

For demonstration and verification purposes,  we  carry  out maximum likelihood estimations with synthetic data. Specifically, $a_{ij} \sim N(r_{ij},\sigma_{ij}^2)$ and $b_i(T)\sim N(b_{i},\sigma_{b_i}^2)$ are sampled from predetermined distributions.  The RMTS(1) can then be  constructed by iteration using equation \ref{eq:RMTS1}. With the time series constructed,  maximum likelihood estimation  procedure will yield the estimated parameters in $a_{ij} \sim N(r_{ij},\sigma_{ij}^2)$ and $b_i(T)\sim N(b_{i},\sigma_{b_i}^2)$. Those estimated parameters can be compared with our initial setups to access estimation accuracy.\\

Table \ref{table:tsparam} lists the parameters that are used to generate the time series. 3 different cases are generated.
For simplicity, we assume all off-diagonal elements have the same distribution and all diagonal elements have another but still the same distribution.

\begin{table}[ht]
	\caption{\textbf{Time series parameters.}} 
	\centering 
	\resizebox{\columnwidth}{!}{\begin{tabular}{c |c| c| c| c| c |c |c| c} 
			\hline\hline 
			Case & Matrix $\mathbf {A}$ dimensions &  $r_{ij}(i\ne j)$ & $\sigma_{ij}(i\ne j)$ &$r_{ij}(i=j)$ & $\sigma_{ij}(i=j)$   & $b_{i}$ & $\sigma_{b_i}$ & T\\  
			\hline 
			1 & 5$\times$5 & 0 & 0.125  & 0 & 0.125& 0&0.5 & 50000\\ 
			2 & 5$\times$5 & 0 & 0.125  & 0 & 0.125& 0&0.5 & 500000\\ 
			3 & 10$\times$10 & 0 & 0.1  & 0.1 & 0.1& 1&1&10000\\
			\hline 
	\end{tabular}}
	\label{table:tsparam} 
\end{table}

Using the synthetic time series generated, we list the estimation results in Table \ref{table:tsout}. All parameters are initialized with one. Nelder–Mead method is used in the optimizations.

\begin{table}[ht]
	\caption{\textbf{Maximum likelihood estimation results.}} 
	\centering 
	\resizebox{\columnwidth}{!}{\begin{tabular}{c |c| c| c| c| c |c |c| c} 
			\hline\hline 
			Case & - log-likelihood &  $r_{ij}(i\ne j)$ & $\sigma_{ij}(i\ne j)$ &$r_{ij}(i=j)$ & $\sigma_{ij}(i=j)$   & $b_{i}$ & $\sigma_{b_i}$ &  iterations\\  
			\hline 
			1 & 192214.4 & 1.382e-4 & 1.239e-1 &  1.115e-3 & -1.360e-1 & -2.010e-3 & 5.010e-1 & 393\\ 
			2 & 1916179.2 & 4.219e-4& -1.286e-1& -1.932e-4& -1.274e-1& 5.142e-4 & 4.991e-1& 495\\ 
			3 & 765322.9 & 1.336e-4&  1.000e-1&  1.025e-1& -9.058e-2&
			0.9973&  1.003 & 576\\
			\hline 
	\end{tabular}}
	\label{table:tsout} 
\end{table}

From this table, we can see that the estimation achieves good accuracy, and the accuracy increases with the number of points(T) in the time series. Practically, the standard deviation estimations might have a negative sign but it won't impact the variances.

\subsection{Statistical properties of $\mathbf {X} (T+1)$}

In this part, we'd like to derive statistical properties for $\mathbf {X} (T+1)$, such as expectations, variance, and covariance.

Assuming each element $a_{ij}(T)$ in  $\mathbf {A(T)}$ is independent of $\mathbf {X} (T) $, and   $a_{ij}(T) \sim N(r_{ij},\sigma_{ij}^2)$,  $b_i(T)\sim N(b_{i},\sigma_{b_i}^2)$. In this paper, we  only consider the situation that those random coefficients have the same distributions over different T(time-invariant).

\subsubsection{Expectation}
To find the relations satisfied by the expectation value. we can start with the iteration relation (Equation \ref{eq:RMTS1}) and apply the law of iterated expectations on this equation.
By the  law of iterated expectations, the expectation values satisfy,

\begin{equation}
	E(\mathbf {X} (T+1))=E(E(\mathbf {X} (T+1)|\mathbf {X} (T)))=E(\mathbf{R}(\mathbf {X} (T))+\mathbf {b(T)})=\mathbf{R}E(\mathbf {X} (T))+\mathbf {b}.
\end{equation}

$\mathbf {R}$ is the matrix representation of $r_{ij}$, and $\mathbf {b}$ is the vector expectation of $b(T)$. Notice that the expectation value of the random matrix $\mathbf {A(T)}$ is defined as $\mathbf{R}$.\\

If the iteration converges, it is easy to see that the converged expectation $\mathbf{e}$ is

\begin{equation*}
	\mathbf{e}=(\mathbf{I}-\mathbf{R})^{-1}\mathbf {b}.
\end{equation*}

Subtracting equation $\mathbf{e}=\mathbf{R}\mathbf{e}+\mathbf {b}$ from the equation of iterated expectation, we have

\begin{equation*}
	E(\mathbf {X} (T+1))-(\mathbf{I}-\mathbf{R})^{-1}\mathbf {b}=\mathbf{R}(E(\mathbf {X} (T)-(\mathbf{I}-\mathbf{R})^{-1}\mathbf {b}).
\end{equation*}

So

\begin{equation}
	E(\mathbf {X} (T))-(\mathbf{I}-\mathbf{R})^{-1}\mathbf {b}=\mathbf{R}^T(E(\mathbf {X} (0))-(\mathbf{I}-\mathbf{R})^{-1}\mathbf {b}).
\end{equation}

Here, $\mathbf{R}^T$ is the product of T(an integer) $\mathbf{R}$ matrices, not transpose.

Assuming $\mathbf {R}$ has eigendecomposition 

\begin{equation}
	\mathbf {R} =\mathbf {Q} \mathbf {\Lambda}\mathbf {Q^{-1}}, 
\end{equation}

$\mathbf {\Lambda}$ is the eigenvalue matrix with eigenvalue $\lambda_i$. Then it is easy to conclude that a necessary condition for convergence is $|\lambda_i|<1$.

\subsubsection{Variance}

Similar to expectation, we can derive properties of the variance.

By the  law of iterated variance, we have

\begin{equation} \label{itv}
	Var(\mathbf {X} (T+1))=E(Var(\mathbf {X} (T+1)|\mathbf {X} (T)))+Var(E(\mathbf {X} (T+1)|\mathbf {X} (T))),
\end{equation}

specifically,

\begin{equation*}
	Var(E(\mathbf {X} (T+1)|\mathbf {X} (T)))=Var(\mathbf{R}\mathbf {X} (T)+\mathbf {b})=\mathbf{R^{\circ2}}Var(\mathbf {X} (T)),
\end{equation*}

and

\begin{equation*}
	E(Var( {X_i} (T+1)|\mathbf {X} (T)))
	=E(\sum_j\sigma_{ij}^2X_j(T)^2+\sigma_{b_i}^2)=\sum_j\sigma_{ij}^2E(X_j(T)^2)+\sigma_{b_i}^2.
\end{equation*}

In matrix notation, define $\mathbf {\Sigma^{\circ 2}}$(elementwise square) as the matrix representation of $\sigma_{ij}^2$, and similarly for $\mathbf{R^{\circ2}}$.

Plug into Equation \ref{itv}, we have

\begin{equation*}
	Var(\mathbf {X} (T+1))=\mathbf {\Sigma^{\circ 2}}E(\mathbf {X^2} (T))+\mathbf {\Sigma_{b}^2}+\mathbf{R^{\circ2}}Var(\mathbf {X} (T)),
\end{equation*}

which can also be expressed as

\begin{equation*}
	Var(\mathbf {X} (T+1))=\mathbf {\Sigma_{b}^2}+(\mathbf {\Sigma^{\circ 2}}+\mathbf{R^{\circ2}})Var(\mathbf {X} (T))+\mathbf{\Sigma^{\circ 2}}(E(\mathbf {X} (T)))^2.
\end{equation*}

When the expectation value converges, we have

\begin{equation}
	Var(\mathbf {X}(T+1))
	=(\mathbf {\Sigma^{\circ 2}}+\mathbf{R^{\circ2}})Var(\mathbf {X} (T))+\mathbf {\Sigma_{b}^2}+\mathbf{\Sigma^{\circ 2}}((\mathbf{I}-\mathbf{R})^{-1}\mathbf {b})^{\circ 2}.
\end{equation}\\

Similar to the expectation, by requiring the iterated  variance  to be convergent, we have

\begin{equation}
	Var(\mathbf {X^*})=(\mathbf{I}-\mathbf {\Sigma^{\circ 2}}-\mathbf{R^{\circ2}})^{-1}(\mathbf {\Sigma_{b}^2}+\mathbf{\Sigma^{\circ 2}}((\mathbf{I}-\mathbf{R})^{-1}\mathbf {b})^{\circ 2}).
\end{equation}

A necessary condition for convergence is: the eigenvalues of $\mathbf {\Sigma^{\circ 2}}+\mathbf{R^{\circ2}}$ are less than 1.

\subsubsection{Covariance}

By the  law of iterated covariance, we have

\begin{equation}
	\begin{split}
		Cov( {X_i}(T+1), {X_j}(T+1))=E(Cov( {X_i}(T+1), {X_j}(T+1)|\mathbf {X} (T)))\\+Cov(E({X_i}(T+1)|\mathbf {X} (T)),E({X_j}(T+1)|\mathbf {X} (T))).
	\end{split}
\end{equation}

When $i\ne j$,

\begin{equation*}
	E(Cov( {X_i}(T+1), {X_j}(T+1)|\mathbf {X} (T)))=E(Cov( \sum_k a_{ik}(T){X_k}(T), \sum_l a_{jl}(T){X_l}(T)|\mathbf {X} (T))).
\end{equation*}

With

\begin{equation*}
	E({X_i}(T+1)|\mathbf {X} (T))=\sum_k r_{ik}X_k(T)+b_i,
\end{equation*}

we have
\begin{equation*}
	Cov(E({X_i}(T+1)|\mathbf {X} (T)),E({X_j}(T+1)|\mathbf {X} (T)))=\sum_{kl}r_{ik}r_{jl}Cov(X_k(T),X_l(T)).
\end{equation*}

So

\begin{equation}\label{eq:cov}
	\begin{split}
		Cov( {X_i}(T+1), {X_j}(T+1))=E(Cov( \sum_k a_{ik}(T){X_k}(T), \sum_l a_{jl}(T){X_l}(T)|\mathbf {X} (T)))\\+\sum_{kl}r_{ik}r_{jl}Cov(X_k(T),X_l(T)).
	\end{split}
\end{equation}
Normally the first term will not be zero, unless $a_{ik}(T)$ and $a_{jl}(T)$ are independent. In GOE and GUE, the symmetric constraints will impact the iteration relation above. We will discuss those cases in the examples below.

\subsection{Examples}

\subsubsection{RMTS(1) from GOE}

The Gaussian Orthogonal Ensemble (GOE) is random symmetric matrices whose upper diagonal entries are independent. The diagonal elements $A_{ii}\sim N(0,2)$. And  off-diagonal elements satisfies  $A_{ij}\sim N(0,1) ~with~  (i<j)$ and $A_{ij}=A_{ji}$. \\

In this case $\mathbf {R}=0$ and $E(\mathbf {X} (T+1))=\mathbf {b}$. 

The variance satisfies  

\begin{equation}
	Var(\mathbf {X} (T+1))=\mathbf {\Sigma_{b}^2}+\mathbf{\Sigma^{\circ 2}}\mathbf {b}^2+\mathbf {\Sigma^{\circ 2}}Var(\mathbf {X} (T)).
\end{equation}

The symmetric constraint plays no role in the calculation of variance since all elements in the same roll of a random matrix are independent.

For GOE, with $\mathbf{Ones}$ defined as matrix with all elements equals one, we have
\begin{equation*}
	\mathbf{\Sigma^{\circ 2}}=\mathbf{Ones}+\mathbf{I}.
\end{equation*}

Notice that under this parameter, the variance will not be stable.

But if we multiply a factor, f to $a_{ij}$, the deviation matrix becomes $f\mathbf{\Sigma}$

The variance may converge at
\begin{equation}
	Var(\mathbf {X_{\infty}})=(\mathbf{I}-f^2\mathbf{\Sigma^{\circ 2}})^{-1}(\mathbf {\Sigma_{b}^2}+f^2\mathbf{\Sigma^{\circ 2}}\mathbf {b}^2).
\end{equation}

For GOE, $a_{ij}=a_{ji}$, so Equation \ref{eq:cov} has non-vanishing terms when $k=j$ and $l=i$. 

\begin{equation*}
	\begin{split}
		Cov( {X_i}(T+1), {X_j}(T+1))=E(Cov( a_{ij}(T){X_j}(T),  a_{ji}(T){X_i}(T)|\mathbf {X} (T)))\\=Var(a_{ij}(T))E(X_i(T)X_j(T)),
	\end{split}
\end{equation*}

which equals

\begin{equation*}
	Cov( {X_i}(T+1), {X_j}(T+1))=\sigma_{ij}^2E(X_i(T)X_j(T))=\sigma_{ij}^2(E(X_i(T)E(X_j(T))+	Cov( {X_i}(T), {X_j}(T))).
\end{equation*}

So
\begin{equation}
	Cov( {X_i}(T+1), {X_j}(T+1))=\sigma_{ij}^2(b_ib_j+	Cov( {X_i}(T), {X_j}(T)).
\end{equation}

This covariance is stable or converges 
when $\sigma_{ij}^2<1$(after multiplying a constant factor on the random matrix).\\

For verification purposes,  we carried out simulations with synthetic time series. Specifically, $a_{ij} $ and $b_i(T)$ are sampled from predetermined GOE distributions and normal distributions.  The RMTS(1) can then be constructed by iteration using Equation \ref{eq:RMTS1}. With the time series constructed,  we can compare the simulated parameters with theoretical results.\\

Table \ref{table:goe} lists the comparison between simulation results and theoretical results. 

\begin{table}[ht]
	\caption{\textbf{Comparison: RMTS(1) from GOE}} 
	\centering 
	\resizebox{\columnwidth}{!}{\begin{tabular}{c |c| c| c| c| c |c |c| c} 
			\hline\hline 
			Case & Matrix $\mathbf {A}$ dimensions &  $r_{ij}(i\ne j)$ & $f\sigma_{ij}(i\ne j)$ &$r_{ij}(i=j)$ & $f\sigma_{ij}(i=j)$   & $b_{i}$ & $\sigma_{b_i}$ & T\\  
			\hline 
			1 & 5$\times$5 & 0 & 0.1 & 0 & 0.1& 1 &1 & 50000\\ 
			2 & 5$\times$5 & 0 & 0.25  & 0 & 0.25& 0.5&0.5 &  50000\\ 
			\hline 
			
	\end{tabular}}		
	\resizebox{\columnwidth}{!}{\begin{tabular}{c |c| c| c| c| c |c } 
			
			\hline\hline 
			Case & E(X) &  Var(X) & Cov(X) &E(X) simulation &Var(X) simulation & Cov mean\\  
			\hline 
			1 & 1 & 1.105 & 0.0101 &0.996,0.992,0.999,0.999,0.999& 1.104,1.106,1.096,1.116,1.114& 0.0065\\ 
			\hline 
			2 & 0.5 & 0.477& 0.0167 &0.502,0.506,0.498,0.505,0.497& 0.478,0.484,0.472,0.477,0.483& 0.0143\\ 
			\hline 
			
	\end{tabular}}
	\label{table:goe} 
\end{table}

The results above demonstrate the correctness of our derivations.

\subsubsection{RMTS(1) from GUE}

For complex random variables, the properties of variances and covariances are slightly different due to the modulus in the definition of variance and covariance.

For example, the variance of a complex random variable is defined as

\begin{equation}
	Var(\mathbf {X} (T+1))=E(\mathbf {|X(T)-E(\mathbf {X(T)})|^2} )=E(\mathbf {|X(T)|^2} )-|E(\mathbf {X(T)})|^2 ).
\end{equation}

From Equation \ref{itv}, we have

\begin{equation*}
	Var(E(\mathbf {X} (T+1)|\mathbf {X} (T)))=Var(\mathbf{R}\mathbf {X} (T)+\mathbf {b})=\mathbf{|R|^{\circ2}}Var(\mathbf {X} (T)).
\end{equation*}

With

\begin{equation*}
	E(Var( {X_i} (T+1)|\mathbf {X} (T)))
	=E(\sum_j|\sigma_{ij}|^2|X_j(T)|^2+|\sigma_{b_i}|^2)=\sum_j|\sigma_{ij}|^2E(|X_j(T)|^2)+|\sigma_{b_i}|^2),
\end{equation*}

We have

\begin{equation*}
	Var(\mathbf {X} (T+1))=\mathbf {|\Sigma_{b}|^2}+(\mathbf {|\Sigma|^{\circ 2}}+\mathbf{|R|^{\circ2}})Var(\mathbf {X} (T))+\mathbf{|\Sigma|^{\circ 2}}(|E(\mathbf {X} (T))|)^2,
\end{equation*}

which may be stable at
\begin{equation}
	Var(\mathbf {X^*})=(\mathbf{I}-\mathbf {|\Sigma|^{\circ 2}}-\mathbf{|R|^{\circ2}})^{-1}(\mathbf {|\Sigma_{b}|^2}+\mathbf{|\Sigma|^{\circ 2}}(|(\mathbf{I}-\mathbf{R})^{-1}\mathbf {b}| )^{\circ 2}). 
\end{equation}

The covariance for complex random variables  is defined as

\begin{equation}
	Cov(X_i (T),X_j (T))=E((X_i (T)-E(X_i (T)\overline{(X_j (T)-E(X_j (T))}  ).
\end{equation}

For GUE, $a_{ij}=\overline{a_{ji}}$, Equation \ref{eq:cov} has nonvanishing terms when $k=j$ and $l=i$.  So

\begin{align*}	
	Cov( {X_i}(T+1), {X_j}(T+1)) &= E(Cov( a_{ij}(T){X_j}(T),  a_{ji}(T){X_i}(T)|\mathbf {X} (T))) \\&=Cov(a_{ij}(T),\overline{a_{ij}(T)})E(X_j(T)\overline{X_i(T)}),
\end{align*}

which equals

\begin{equation*}
	Cov( {X_i}(T+1), {X_j}(T+1))=Cov(a_{ij}(T),\overline{a_{ij}(T)})(E(X_j(T)E(\overline{X_i(T)})+	\overline{Cov( {X_i}(T), {X_j}(T))}).
\end{equation*}

$Cov(a_{ij}(T),\overline{a_{ij}(T)})$ is actually the pseudo-variance of $a_{ij}$, and in our special case of GUE, evaluate to $Re(\sigma_{ij})^2-Im(\sigma_{ij})^2$

So 
\begin{equation}
	Cov( {X_i}(T+1), {X_j}(T+1))=(Re(\sigma_{ij})^2-Im(\sigma_{ij})^2)(\overline{b_i}b_j+	\overline{Cov( {X_i}(T), {X_j}(T))})
\end{equation}

The Gaussian Unitary Ensemble (GUE) are random Hermitian matrices whose entries are independent. The diagonal elements $A_{ii}\sim N(0,1)$. And  off-diagonal elements satisfies  $A_{ij}\sim N(0,0.5)+ N(0,0.5)\imath~with~  (i<j)$. In this case $\mathbf {R}=0$ and $E(\mathbf {X} (T+1))=\mathbf {b}$. 

Similar to GOE, a comparison can be carried out to verify those relations.

%
%



\section{N-th order time series with random matrix coefficients, RMTS(n).}
Similar to RMTS(1), we can easily define RMTS(n) by:

\textbf{Defination}: A nth-order vector autoregressive time series with random matrix coefficients, denoted by RMTS(n), is defined by the iteration relation:

\begin{equation} 
	\mathbf {X}(T+1)=\sum_{i=T-n+1}^{T} \mathbf{A}(T_i)\mathbf {X}(T_i))+\mathbf {b}(T).
\end{equation}

This equation can be converted to an RMTS(1) by rewriting this iteration relation in a matrix form.

\begin{equation} 
	{\begin{bmatrix}\mathbf {X} (T+1)\\\mathbf {X} (T)\\\mathbf {...} \\\mathbf {X} (T-n+2)
	\end{bmatrix}}={\begin{bmatrix}\mathbf {A}(T) &  \mathbf {A}(T-1)&\mathbf {...}  &  \mathbf {A}(T-n+1) \\\mathbf {I}  & \mathbf {0} &\mathbf {...}& \mathbf {0}\\\mathbf {...} \\\mathbf {0}  & \mathbf {...} &\mathbf {I}& \mathbf {0} \\\end{bmatrix}}{\begin{bmatrix}\mathbf {X} (T)\\\mathbf {X} (T-1)\\\mathbf {...} \\\mathbf {X} (T-n+1)
	\end{bmatrix}}+{\begin{bmatrix}\mathbf {b} (T)\\\mathbf {0} \\\mathbf {...} \\\mathbf {0} 
	\end{bmatrix}}.
\end{equation}

As a result, the properties for RMTS(1) directly apply to the equation above.
\section{First-order Random Matrix differential equations.}
With the properties of random matrix time series derived, it is natural for us to consider the continuous case for comparison, which is the first-order random matrix differential equations.

A typical form of the first order random matrix differential equations can be written as,

\begin{equation} \label{eq:RMDF1}
	\dfrac{d \mathbf {X} (T)}{d T}=\mathbf{A}(T)\mathbf {X} (T)+\mathbf {b}(T).
\end{equation}

Here, $\mathbf{A(T)}$ and $\mathbf{b(T)}$ have the same definition as in the time series. Notice that when this equation is discretized, it becomes an RMTS(1) with coefficients $\Delta T(\mathbf{A}(T)+\mathbf{I})$ and $\Delta T\mathbf{b}(T)$.

We can find the solution to this equation by taking the limit of the corresponding discretized time series. For example, consider the trivial case which is $\mathbf{b}(T)$=0, separating $T$ into $n$ equally spaced intervals, by iteration, it is easy to see that the solution is,

\begin{equation*}
	\mathbf {X} (T)=\lim_{n \to \infty}{\prod_n(\dfrac{T\mathbf{A}(t)}{n}+\mathbf{I})^n}		 \mathbf {X} (0).
\end{equation*}\

We define the limit as $\mathbf{RMEXP(A,T)}$ if the limit exists, then our solution can be expressed as

\begin{equation*}
	\mathbf {X} (T)=\mathbf {RMEXP(A,T)}\mathbf {X} (0).
\end{equation*}\\

Using this definition in Equation \ref{gs}.   The solution, when  $\mathbf{b}(T) \ne 0$, can be written as

\begin{equation}
	\mathbf {X} (T)=\mathbf {RMEXP(A,T)}(\mathbf {X} (0)+\int_{t=0}^{T} \mathbf {RMEXP(A,t)}^{-1}\mathbf {b} (t)dt).
\end{equation}

Here the integral should be understood in the sense of a summation of a random process. Notice that the solution is similar to the corresponding one in ordinary differential equations.\\

Generically, $RMEXP(A,T)$ can be difficult to compute, yet a few known solutions exist. The simplest example is: A(t) is a 1$\times$1 random scalar and $b(T)=0$. Each scalar is in a normal distribution, written as $a(t)$, of zero mean and unit variance.

In this case, exact solutions had been derived\cite{jackson2002products}. The limit, defined as variable Y,

\begin{equation*}
	RMEXP(a,T)=Y=\lim_{n \to \infty}{\prod_t(\dfrac{Ta(t)}{n}+1)^n},
\end{equation*}

has a probability density function of
\begin{equation*}
	\mathcal{P}(Y)=\frac{1}{Y \sqrt{2 \pi} T} \exp \left[-\frac{(\log Y+T^2 / 2)^{2}}{2 T^2}\right],
\end{equation*}

which is lognormal.

When $b(T)=0$ and $\mathbf{A}(t)$ is a 2$\times$2 matrix, with each element of $\mathbf{A(T)}$ on a normal distribution  of zero mean and unit variance, the random matrix exponential $RMEXP(A,T)$ is still exact solvable. The limiting matrix probability distribution is very complicated, which will be omitted here. Readers can see Ref. \cite{jackson2002products} for an analytical expression.

Normally, statistical properties of $\mathbf {X} (T)$, such as expectation and variance, will diverge when $T$ goes infinity. This is because of the specific form of this differential equation \ref{eq:RMDF1} (Since an identity matrix is added to the to $\mathbf {A} (T)$). 

With the fruitful ongoing research of the products of random matrices, we expect more analytical solutions of such differential equations can be derived, and this topic can be researched systematically.

%
%
%

\section{Applications}
Time series models can be used in obtaining the structure that produced the observed data or fitting a model and using it to make forecasts. Time series models can be used in many applications such as economic forecasting and quality control, etc. In this article, we propose a simple scenario where our random matrix times series models can be used. 

We can interpret vector $\mathbf {X} (T)$ as the number of restaurants of a specific category(for example, traditional, fast food, etc.) in a city at time  $T$. It is easy to imagine that the number of restaurants in one category at the next timestamp depends on the number of restaurants in another category at the current timestamp due to market competition.  It is also natural that the number of restaurants in the category at the next timestamp depends on the number of restaurants in the same category at the current timestamp due to shut down and internal competition. The coefficient matrix $\mathbf {A} (T)$  can be used to capture this dependency. The coefficient  $\mathbf {b} (T)$ in the same category can be interpreted as the number of newly opened restaurants in the same category.

For example,  similar to the maximum likelihood estimation, we demonstrate this process with synthetic data. Table \ref{table:aptsparam} lists the parameters that are used to generate the time series. Those values represent the dependency coefficient in the scenario discussed above.  

\begin{table}[ht]
	\caption{\textbf{Synthetic time series parameters.}} 
	\centering 
	\begin{tabular}{c| c| c| c| c |c |c| c} 
		\hline\hline 
		Matrix $\mathbf {A}$ dimensions &  $r_{ij}(i\ne j)$ & $\sigma_{ij}(i\ne j)$ &$r_{ij}(i=j)$ & $\sigma_{ij}(i=j)$   & $b_{i}$ & $\sigma_{b_i}$ & T\\  
		\hline 
		3$\times$3 & -0.1 & 0.1 & 0.5 & 0.1& 1&0& 50000\\ 
		\hline
	\end{tabular}
	\label{table:aptsparam} 
\end{table}

Using this synthetic time series generated, we list the estimation results in Table \ref{table:aptsout}. All parameters are initialized with one. Powell's method is used in the optimizations.

\begin{table}[!ht]
	\caption{\textbf{Model fitting results with maximum likelihood estimation. (Number of iterations: 31)}} 
	\centering 
	\begin{tabular}{ c| c| c| c  } 
		\hline\hline 
		$\mathbf {R} $ & $\mathbf {\Sigma}$ &$\mathbf {b}$  &$\mathbf {\Sigma_b}$ \\  
		\hline 
		$\begin{bmatrix}0.4974, -0.1014 , -0.08910\\-0.0994,  0.4934,
			-0.0994\\-0.0973, -0.1031,  0.4920
		\end{bmatrix} $&  	
		$\begin{bmatrix}-0.0977,
			-0.0972 -0.0969\\ -0.1003, -0.09684, -0.0974\\
			-0.0958 , -0.1023, -0.0925
		\end{bmatrix} $
		& $\begin{bmatrix}0.989\\ 1.008\\
			1.010\end{bmatrix}$& $\begin{bmatrix}0.0508\\ -0.0410\\ -0.0666\end{bmatrix} $\\ 
		\hline 
	\end{tabular}
	\label{table:aptsout} 
\end{table}

From this table, we can see that the estimation achieves good accuracy and captures the dynamics between categories in the data.

\section{Conclusions}

In this paper, we discussed random matrix times series, with properties numerically verified. This approach is novel and it is an interesting question and a promising direction to explore more structures with random matrices and identify potential applications.

\section*{Acknowledgements}

Python programs used for the simulations in this article are available on the Authors' GitHub page. No potential conflict of interest was reported by the author(s). This research received no specific grant from any funding agency in the public, commercial, or not-for-profit sectors.

\end{document}